\documentclass[aps,prb,twocolumn,showpacs,groupedaddress]{revtex4-1}
\usepackage{amsmath}
\usepackage{amssymb}
\usepackage{graphicx}
\usepackage{dcolumn}
\usepackage{bm}
\usepackage{subfigure}

\newcommand{\hd}{H$\downarrow$}

\usepackage[normalem]{ulem}
\usepackage{color}

\begin{document}

\title{Spin-polarized hydrogen adsorbed on the surface of 
superfluid $^{\bf 4}$He}
\author{J. M. Mar\'\i n$^{\rm a}$, L. Vranje\v{s} Marki\'c$^{\rm b}$ 
and J. Boronat$^{\rm a}$}
\affiliation{
$^{\rm a}$ Departament de F\'{i}sica i Enginyeria Nuclear, Universitat Polit\`{e}cnica 
de Catalunya, Campus Nord B4-B5, E-08034, Barcelona, Spain \\
$^{\rm b}$  Faculty of Science, University of Split, HR-21000 Split,
Croatia
} 

\begin{abstract}
The experimental realization of a thin layer of spin-polarized hydrogen \hd
\  adsorbed on top of the surface of superfluid $^4$He provides one of the
best examples of a stable nearly two-dimensional quantum Bose gas. We
report a theoretical study of this system using quantum Monte Carlo methods
in the limit of zero temperature. Using the full Hamiltonian of the system,
composed of a superfluid $^4$He slab and the adsorbed \hd \ layer, we
calculate the main properties of its ground state using accurate models for
the pair interatomic potentials. Comparing the results for the layer with
the ones obtained for a strictly two-dimensional (2D) setup, we analyze the
departure from the 2D character when the density increases. Only when the
coverage is rather small the use of a purely 2D model is justified. The
condensate fraction of the layer is significantly larger than in 2D at the
same surface density, being as large as 60\% at the largest coverage
studied. 



\end{abstract}

\maketitle

\section{Introduction}

Electron-spin-polarized hydrogen (\hd) was proposed long time ago as  
the system in which a Bose-Einstein condensate state (BEC) could be
obtained.~\cite{stwaley,miller} 
Intensive theoretical and
experimental work was made in the eighties and nineties of the past century
to devise experimental setups able to reach the predicted density and
temperature regimes for BEC.~\cite{silvera1,greytak,silvera2} 
The high recombination rate in the walls of
the containers hindered this achievement for a long time, and only after
working with a wall-free confinement, Fried \textit{et al.}~\cite{fried} were able to
realize its BEC in 1998. However, this was not the first BEC  
because three years before the BEC state was impressively obtained working
with cold  metastable alkali gases.~\cite{becgas} The same year BEC of \hd \ was obtained,         
Safonov \textit{et al.}~\cite{safonov} observed for the first time a quasi-condensate of
nearly two-dimensional \hd \ adsorbed on the surface of superfluid $^4$He. 

In spite  of hydrogen losing the race against alkali gases to be the first
BEC system, it still deserves interest for both theory and experiment.
Hydrogen is the lightest and most abundant element of the Universe and, when
it is spin polarized with the use of a proper magnetic field, it is the only
system that remains in the gas state down to the limit of zero temperature. \hd
\ is therefore extremely quantum matter. A standard measure of the
quantum nature of a system is the de Boer parameter~\cite{miller}
\begin{equation}
\eta = \frac{\hbar^2}{m \epsilon \sigma^2} \ ,
\label{deboer}
\end{equation}
with $\epsilon$ and $\sigma$ the well depth and core radius of the pair interaction,
respectively. According to this definition, $\eta=0.5$ for \hd \ 
which is the largest value for $\eta$ among all the quantum fluids 
(for instance, $\eta=0.2$ for $^4$He). This large value for $\eta$ results
from the shallow minimum ($\sim 6$ K) of the triplet potential $b$~$^3\Sigma_u^+$ 
between spin-polarized hydrogen atoms and their small mass.~\cite{kolos}

Adsorption of \hd \ on the surface of liquid $^4$He has been extensively
used because of its optimal
properties.~\cite{walraven,berkhout,mosk,ahokas,ahokas2,jarvinen} On one hand, 
the interaction of any
adsorbant with the $^4$He surface is the smallest known, and on the other,
at temperatures $T < 300$ mK the $^4$He vapor pressure is negligible and
thereby above the free surface one can reasonably assume vacuum. In fact,
liquid $^4$He was also extensively used in the search of the 
three-dimensional \hd \ BEC state when the cells were coated with helium films to
avoid adsorption of \hd \ on the walls and the subsequent recombination to
form molecular hydrogen H$_2$.~\cite{silvera1,greytak,silvera2} 
Helium is chemically inert and only a small
fraction of $^3$He ($6.6$ \%) is soluble in bulk $^4$He; spin-polarized
hydrogen, and its isotopes deuterium and tritium, are expelled to the
surface where they have a single bound state. For instance, in the case of
\hd, the chemical potential of a single atom in bulk $^4$He
is~\cite{marin1} $36$ K to be
compared with the negative value on the surface, $-1.14$
K.~\cite{safonov2,mantz,silvera3}

The quantum degeneracy of \hd \ adsorbed on $^4$He is quantified by
defining the quantum parameter $\sigma \Lambda^2$, with $\sigma$ the
surface density and $\Lambda$ the thermal de Broglie wave
length.~\cite{jarvinen}
Experiments try to increase this parameter as much as possible by
increasing the surface density and lowering the temperature of the film.
To this end, two methods for local compression have been used. The first
one, that relies on the application of a high magnetic field, is able to
attain large  quantum parameter values, $\sigma \Lambda^2 \simeq
9$.~\cite{safonov,mosk}
However, to measure the main properties of the quasi-two dimensional gas 
becomes difficult due to the large magnetic field.~\cite{jarvinen2} 
An alternative to this
method is to work with thermal compression, in which a small spot on the sample
cell is cooled down to a temperature below the one of the
cell.~\cite{matsubara,vasyliev}  This second
method achieves lower values for quantum degeneracy  $\sigma \Lambda^2
\simeq 1.5$ but allows for direct observation of the sample. Up to now, it
has not been possible to arrive to the value $\sigma \Lambda^2
\simeq 4$ where the   Berezinskii-Kosterlitz-Thouless superfluid transition
is expected to set in. Nevertheless, the quantum degeneracy of the gas has been
observed as a decrease of the three-body recombination rate at temperatures
$T=120$-$200$ mK and densities $\sigma \simeq 4 \times 10^{12}$
cm$^{-2}$.~\cite{jarvinen2}

The zero-temperature equations of state of bulk gas~\cite{leandra1} \hd \
and liquid~\cite{leandra2}
T$\downarrow$ have been recently calculated using accurate quantum Monte
Carlo methods. Properties like the condensate fraction, distribution
functions and localization of the gas(liquid)-solid phase transitions have
been established with the help of the ab initio \hd-\hd  \ interatomic
potential.~\cite{kolos,jamieson,yan} From the theoretical side, much less is 
known about the ground-state
properties of two-dimensional \hd \ or \hd  \ adsorbed on a free $^4$He surface. 
In a pioneering work, Mantz and Edwards~\cite{mantz} used the variational Feynman-Lekner
approximation to calculate the effective potential felt by a hydrogen atom
on the $^4$He surface. Solving the Schr\"odinger equation for the atom in
this effective potential they concluded that \hd, D$\downarrow$, and
T$\downarrow$ have a single bound state and calculated the respective
binding energies. The main drawback of this treatment is that the adsorbent
is substituted by an effective field representing a static and undisturbed
surface. In fact, a quantitatively accurate approach to this problem
requires a good model for the $^4$He surface.~\cite{krotscheck} The use of accurate He-He
potentials and ground-state quantum Monte Carlo methods has proved to be
able to reproduce experimental data directly related to the surface, like the surface
tension and the surface width.~\cite{marin2} In the present work, we rely
on a similar
methodology to the one previously  used in the study of the free $^4$He
surface~\cite{marin2} in order 
to microscopically characterize the ground-state of \hd \ adsorbed on its
surface. Our study is complemented by a purely two-dimensional simulation
of \hd \ in order to establish the degree of two-dimensionality of the
adsorbed film.

The rest of the paper is organized as follows. The quantum Monte Carlo
method used for this study is described in Sec. II. The results obtained
for \hd \ adsorbed on the $^4$He surface within a slab geometry are
presented in Sec. III together with the comparison with the strictly
two-dimensional case. Finally, Sec. IV comprises a brief summary and the
main conclusions of the work.

\section{Quantum Monte Carlo method}

We have studied the ground-state (zero temperature) properties of a thin
layer of \hd \ adsorbed on the free surface of a $^4$He slab and also the
limiting case of a strictly two-dimensional (2D) \hd \ gas. Focusing first
on the slab geometry, the Hamiltonian of the system composed by $N_{\rm
He}$ $^4$He and $N_{\rm H}$  \hd \  atoms is
\begin{eqnarray}
H & =  & -\frac{\hbar^2}{2 m_{\rm He}} \sum_{I=1}^{N_{\rm He}} {\bm \nabla}_I^2
-\frac{\hbar^2}{2 m_{\rm H}} \sum_{i=1}^{N_{\rm H}} {\bm \nabla}_i^2
+ \sum_{1=I<J}^{N_{\rm He}} V_{\rm He-He} (r_{IJ})  \nonumber \\   
 & & + \sum_{1=i<j}^{N_{\rm H}} V_{\rm H-H} (r_{ij})
+ \sum_{1=I,i}^{N_{\rm He},N_{\rm H}} V_{\rm He-H} (r_{Ii})  \ , 
\label{hamiltonian}
\end{eqnarray}   
with capital and normal indices standing for $^4$He and \hd \ atoms,
respectively. The pair potential between He atoms is the Aziz HFD-B(HE)
model~\cite{aziz} used extensively in microscopic studies of liquid and solid helium.   
The \hd-\hd \ interaction ($b~^3\Sigma_u^+$ triplet potential) was calculated
with high accuracy by Kolos and Wolniewicz (KW).~\cite{kolos} More recently, this potential
has been recalculated up to larger interatomic distances by
Jamieson, Dalgarno, and Wolniewicz (JDW).~\cite{jamieson} We have used the JDW data  
smoothly connected
with the long-range behavior of the \hd-\hd \ potential as calculated by Yan
\textit{et al.}~\cite{yan} The JDW potential has a core diameter of $3.67$
\AA\, and a minimum $\epsilon=-6.49$ K (slightly deeper than KW) 
at a distance $r_{\text m}=4.14$ \AA. Finally, we take the H-He pair
potential from Das \textit{et al.};~\cite{das} this model has been
used in the past in the study of a single \hd \ impurity~\cite{marin1} in liquid $^4$He
and in mixed T$\downarrow$-$^4$He clusters.~\cite{petar} The Das
potential~\cite{das} has a 
minimum $\epsilon=-6.53$ K at a distance  $r_{\text m}=3.60$ \AA.      

The quantum $N$-body problem is solved stochastically using the diffusion
Monte Carlo (DMC) method.~\cite{hammond} DMC is nowadays one of the most accurate 
tools for the study of
quantum fluids and gases, providing exact results for boson systems within
some statistical errors. In brief, DMC solves the imaginary-time ($\tau$) $N$-body 
Schr\"odinger equation for the  function $f({\bm R},\tau)=\psi({\bm R})
\Psi({\bm R},\tau)$, with  $\Psi_0({\bm R}) =\lim_{\tau \to \infty} \Psi({\bm
R},\tau)$ the exact ground-state wave function. The auxiliary wave function
$\psi({\bm R})$ acts as a guiding wave function in the diffusion process
towards the ground state when $\tau \to \infty$. The direct statistical
sampling  with  $f({\bm R},\tau)$, called mixed estimator, is unbiased for
the energy but not completely for operators which do not commute with the
Hamiltonian. In these cases, we rely on the use of pure estimators based on
the forward walking strategy.~\cite{pures} The influence of the finite time step used in
the iterative process is reduced by working with a second-order expansion
for the imaginary-time Green's function.~\cite{dmccasu} The last systematic error that one
has to deal with is the finite number of walkers ${\bm R}_i$ which
represent the wave function $\Psi({\bm R},\tau)$. As usual, we analyze
which is the number of walkers required to reduce any bias coming from it
to the level of the statistical uncertainties.

The $^4$He surface is simulated using a slab which grows symmetrically in
the $z$ direction and with periodic boundary conditions in the $x-y$
plane.~\cite{marin2}
The guiding wave function is then the product of two terms
\begin{equation}
\psi ({\bm R}) = \psi_J ({\bm R}) \, \phi({\bm R})  \ ,
\label{trialslab}
\end{equation}  
the first one accounting for dynamical correlations induced by the
interatomic potentials and the second for the finite size of the liquid in
the $z$ direction. Explicitly, $\psi_J ({\bm R})$ is built as a product of
two-body Jastrow factors between the different particles,
\begin{equation}
\psi_J ({\bm R}) = \prod_{1=I<J}^{N_{\rm He}} f_{\rm He}(r_{IJ}) 
\prod_{1=i<j}^{N_{\rm H}} f_{\rm H}(r_{ij})
\prod_{1=I,i}^{N_{\rm He},N_{\rm H}} f_{\rm He-H}(r_{Ii}) \ .	
\label{jastrow}
\end{equation}
The one-body correlations that confine the system to a slab geometry are
introduced in $\phi({\bm R})$,
\begin{equation}
\phi({\bm R}) = \prod_{I=1}^{N_{\rm He}} h_{\rm He}(z_I) 
\prod_{i=1}^{N_{\rm H}} h_{\rm H}(z_i)  \ . 
\label{hslab}
\end{equation}
The $^4$He-$^4$He ($f_{\rm He}(r)$) and $^4$He-\hd \ ($f_{\rm He-H}(r)$) two-body 
correlation factors (\ref{jastrow}) are chosen of Schiff-Verlet type,
\begin{equation}
f(r) = \exp \left[ -\frac{1}{2} \left( \frac{c}{r} \right)^5 \right] \ ,
\label{sverlet}
\end{equation} 
whereas the \hd-\hd \ one is taken as
\begin{equation}
 f_{\rm H}(r) = \exp [ -b_1 \exp(-b_2 r)] \ ,
 \label{fhydro}
\end{equation}
because it has been shown to be variationally better for describing the
hydrogen correlations.~\cite{leandra1} The parameters entering Eqs.
(\ref{sverlet},\ref{fhydro}) have been optimized using the variational
Monte Carlo method. We have used $c_{\rm He}=c_{{\rm He-H}}= 3.07$ \AA,
$b_1=101$, and $b_2=1.30$ \AA$^{-1}$, neglecting their slight dependence on
density. The one-body functions in Eq. (\ref{hslab}) are of Fermi type,
\begin{equation}
h(z)= \left\{ 1 + \exp [ \, k ( \, |z-z_{\text{cm}}| - z_0)  ] \right\}^{-1}   \ ,
\label{fermi}
\end{equation}
with variational parameters $k$ and $z_0$ related to the width and location of the
interface, respectively. The main goal of these one-body terms is to avoid
eventual evaporation of particles by introducing a restoring drift force only
when particles want to escape to unreasonable distances. 
Any spurious kinetic energy contribution due 
to the movement of the center of mass of the full system ($^4$He+\hd) 
is removed by subtracting $z_{\text{cm}}$
from each particle coordinate $z$, either of $^4$He or \hd, 
in Eq. (\ref{fermi}). The optimal values
used in the DMC simulations are $z_0$($^4$He)$=22.10$ \AA, 
$z_0$(\hd)$=37.06$ \AA, and $k$($^4$He)$=k$(\hd)$=1$ \AA$^{-1}$.

Our study of the thin layer of \hd \ adsorbed on $^4$He is complemented
with some calculations of a strictly 2D \hd \ gas with the Hamiltonian
\begin{equation}
H_{\text{2D}}= 
-\frac{\hbar^2}{2 m_{\rm H}} \sum_{i=1}^{N_{\rm H}} {\bm \nabla}_i^2
+ \sum_{1=i<j}^{N_{\rm H}} V_{\rm H-H} (r_{ij}) \ , 
\label{hamiltonian2d}
\end{equation}  
using as a guiding wave function a Jastrow factor with the same two-body
correlation factors as in the slab (\ref{fhydro}).~\cite{h2d}

\section{Results}

The $^4$He surface where \hd \ is adsorbed is simulated with the DMC method using a
slab geometry. We use a square cell in the $x-y$ plane that is made
continuous by considering periodic boundary conditions in both directions.
In the transverse  direction $z$ the system is finite, with two symmetric
free surfaces at the same distance from the center $z=0$. The surface of the
basic simulation cell is $A=290.30$ \AA$^2$ and $N_{\rm He}=324$. With these
conditions we guarantee an accurate model for the free surface of $^4$He, as
shown in Ref. \onlinecite{marin2}.

\begin{figure}
\centerline{
\includegraphics[width=0.9\linewidth]{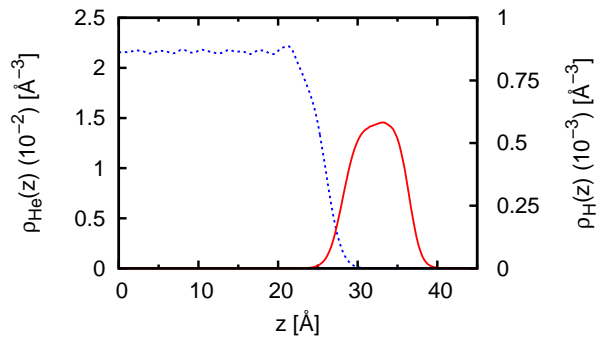}}%
\caption{(Color online) Density profile of the $^4$He slab (dashed line)
and of the \hd \ adsorbed gas (solid line) corresponding to a surface
density  $\sigma=9.57 \times 10^{-3}$ \AA$^{-2}$.       }  
\label{fig1}
\end{figure}

On top of one of the slab surfaces we introduce a variable number $N_{\rm
H}$  of 
\hd \ atoms that form a thin layer of surface densities $\sigma= N_{\rm H}/A$.
In order to reach lower densities than $\sigma=1/A$ we have replicated the basic
slab cell the required number of times. In Fig. \ref{fig1}, we show the
density profiles of the $^4$He slab and of the \hd \ layer for a surface
density $\sigma=9.57 \times 10^{-3}$ \AA$^{-2}$. This layer has an
approximate width of 8 \AA \ and virtually \textit{floats} on the helium
surface: the center of the \hd \ layer is located out of the surface, where
the $^4$He density is extremely small. The picture is similar to the one
obtained previously by Mantz and Edwards~\cite{mantz} in a variational description 
of the adsorption of a single \hd \ atom. However, 
contrarily to the exponential tail of
the density profile derived by Krotschek and Zillich~\cite{krotscheck} in
a thorough description of the impurity problem, we observe a faster decay to
zero and a rather isotropic profile. We attribute this difference to the residual bias of the one-body factor $h(z)$ (\ref{fermi}) 
used to avoid spurious evaporation of particles. 
On the other hand, the more well studied case of $^3$He
adsorbed on the $^4$He surface shows a similar density
profile,~\cite{guardiola} located on
the surface, but in this case centered not so far from the bulk.

\begin{figure}
\centerline{
\includegraphics[width=0.85\linewidth]{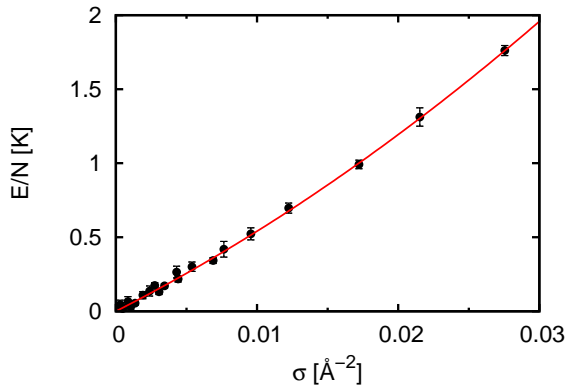}}%
\caption{(Color online) Energy per particle of \hd \ on top of the $^4$He surface
(points with error bars). The energy at the zero-dilution limit is subtracted in such
a way that the energy is zero in the limit $\sigma \rightarrow 0$. The line on top of
the DMC data corresponds to the polynomial fit of Eq. (\ref{fitslab}). }  
\label{fig2}
\end{figure}

\begin{figure}[b]
\centerline{
\includegraphics[width=0.85\linewidth]{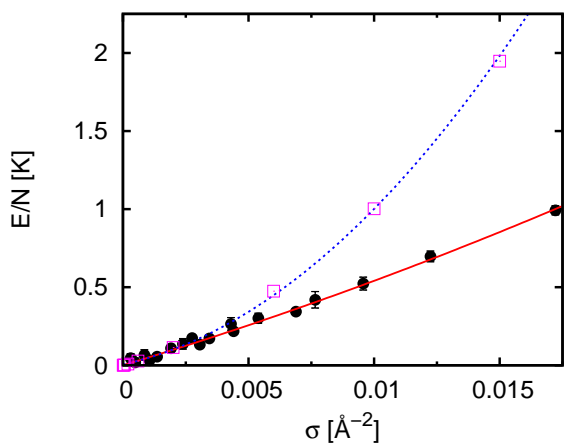}}%
\caption{(Color online) Comparison between the energy per particle of \hd \ adsorbed
on the $^4$He slab (full circles) and the energy of purely two-dimensional \hd \ (open
squares). The solid line is the polynomial fit (\ref{fitslab}) and the dashed line is
a fit of the 2D energies (\ref{fit2d}). }  
\label{fig3}
\end{figure}

One of the most relevant magnitudes that characterize the \hd \ film is its
energy per particle at different coverages. In Fig. \ref{fig2}, we plot the
DMC energy per particle of \hd \ as a function of the surface density
$\sigma$. In order to better visualize the energy of the adsorbed gas, we have
subtracted from the computed energies the energy in the infinite dilution 
limit $\sigma
\rightarrow 0$. The energy increases monotonously with the density and its
behavior is well accounted for by the simple polynomial law   
\begin{equation}
E/N (\sigma)=B \sigma + C \sigma^2 \ ,
\label{fitslab}
\end{equation}
with optimal parameters $B = 48(2)$ K\AA$^2$ and $C = 5.6(9) \times 10^2$ K\AA$^4$,
the figures in parenthesis being the statistical uncertainties.

\hd \ floating on top of the $^4$He free surface has been currently
considered as a nice representation of a quasi-two-dimensional quantum gas.
In order to be quantitatively accurate in this comparison, we have carried
out DMC simulations of strictly 2D \hd \ gas without any adsorbing
surface.~\cite{h2d}
The results obtained for the energy per particle of the 2D gas at different densities are
shown in Fig. \ref{fig3}. The energies are well reproduced by a polynomial
law 
\begin{equation}
E/N (\sigma)=B_{\rm{2D}} \sigma + C_{\rm{2D}} \sigma^2 \ ,
\label{fit2d}
\end{equation}
with $ B_{\rm{2D}}= 35(3)$ K\AA$^2$  and $ C_{\rm{2D}}= 6.4(1) \times 10^4$
K\AA$^4$. In the same figure, we plot the energies for the adsorbed gas at
the same coverage. As one can see, the agreement between the strictly 2D
gas and the film is good for densities $\sigma \alt 5 \times 10^{-3}$
\AA$^{-2}$. At higher densities, the additional degree of freedom in the $z$ direction 
makes  the growth of the energy with the surface density  in the layer
 nearly linear up to the shown density, in contrast with the significant
quadratic increase observed in the 2D gas ($C << C_{\rm{2D}}$).

\begin{figure}
\centerline{
\includegraphics[width=0.85\linewidth]{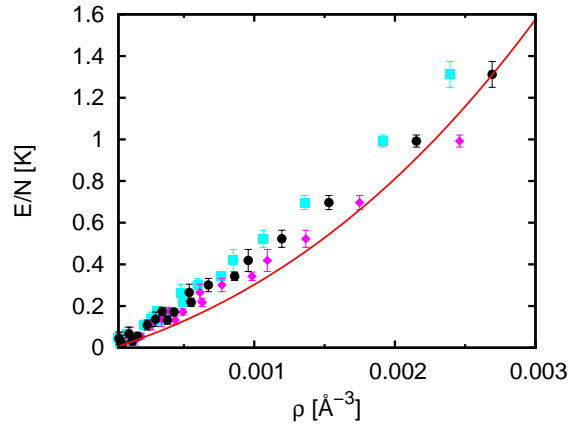}}%
\caption{(Color online) Comparison between the energy per particle of \hd \ adsorbed
on the $^4$He slab and the energy of bulk \hd \ (solid line) from Ref. \onlinecite{leandra1}. 
Full squares, full circles, and full diamonds correspond to the layer where we have considered a 
width in $z$ of 9, 8, and 7 \AA, respectively.  }  
\label{fig4}
\end{figure}

A possible scenario when the density increases and the equation of 
state of the layer departs from the 2D law is the existence of a nearly
three-dimensional (3D) gas. We have analyzed this possibility by considering a 
width in $z$ given by the density profile (Fig. \ref{fig1}) and by estimating the 3D
density of the adsorbed gas as the coverage divided by the layer width. In Fig.
\ref{fig4}, we show the energy per particle of adsorbed \hd \ as a function of
the density considering our best estimation for the layer width, 
$z=8$ \AA, and also $z=9$ and 7 \AA. The possible 3D behavior of the energy 
is analyzed by comparing the results of the layer with the ones of the bulk 3D gas. 
At low densities, the energies of the adsorbed phase are higher than the 3D gas and, when the
density increases, both results tend to cross. As one can see,
the energies of adsorbed \hd \ are not well described by a 3D equation of
state at any density within the regime studied.

\begin{figure}
\centerline{
\includegraphics[width=0.55\linewidth,angle=-90]{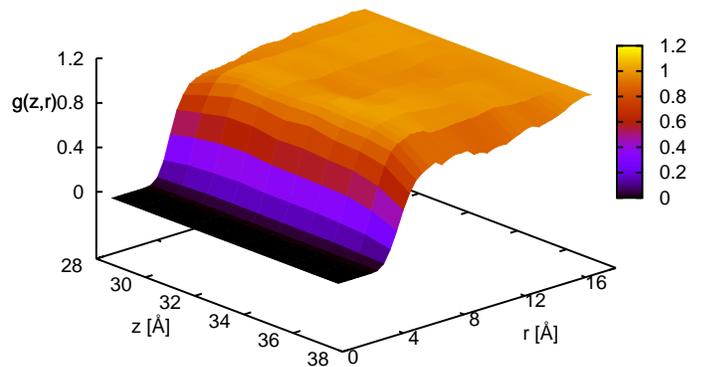}}%
\caption{(Color online) Two-body distribution function $g(z,r)$ of \hd \ adsorbed on
$^4$He, with $r=\sqrt{x^2+y^2}$, at surface density $\sigma=0.0215$ \AA$^{-2}$.   }  
\label{fig5}
\end{figure}

The structure and the distribution functions of \hd \ atoms in the layer can be
studied by doing slices of small width ($\Delta z = 1$ \AA) and, within a given slice, 
as a function of the radial distance between particles in the plane $r=\sqrt{x^2+y^2}$. In
Fig. \ref{fig5}, we report results of the two-body radial distribution
function $g(z,r)$ where $z$ is the distance to the center of the $^4$He
slab at a coverage $\sigma=0.0215$ \AA$^{-2}$. Around the center of the \hd
\ density profile, $g(r)$ is nearly independent of $z$ with a main peak of a
height smaller than 1.2. In the wings of $\rho_{\text H}(z)$, where the
local density is smaller, $g(r)$ shows less structure and the noise of the
DMC data also increases due to low statistics.

\begin{figure}
\centerline{
\includegraphics[width=0.85\linewidth]{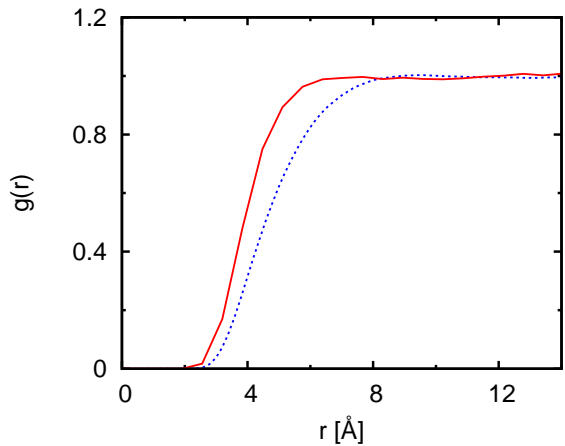}}%
\caption{(Color online) Comparison between the two-body distribution function in the
center of the slab, corresponding to the density $\sigma=0.0095$ \AA$^{-2}$ (solid
line) with the one corresponding to a purely 2D \hd \ gas at the same surface density
(dotted line).  }  
\label{fig6}
\end{figure}

It is interesting to know if the spatial structure of \hd \ atoms on the
$^4$He surface is similar to the one in a strictly 2D geometry. To this
end, we show in Fig. \ref{fig6} results of the radial distribution function
for both systems at the same surface density ($\sigma=0.0095$ \AA$^{-2}$).
The result corresponding to the layer is taken from a slice $\Delta z$ in
the center of the density profile. As one can see, both functions do not
show any significant peak because the density is rather small. However, the
behavior at small interparticle distances is appreciably different. In the
layer, atoms can be closer (in the in-plane distance $r=\sqrt{x^2+y^2}$ ) 
than in 2D because of the small but nonzero
width of the slice used for its calculation. In fact, we have shown
previously in Fig.
\ref{fig2} that, at the density  $\sigma=0.0095$ \AA$^{-2}$ used in Fig.
\ref{fig6}, the energies per particle of the layer and the strictly 2D gas
start to be significantly different, in agreement with the differences
observed here in the distribution function $g(r)$.

\begin{figure}
\centerline{
\includegraphics[width=0.55\linewidth,angle=-90]{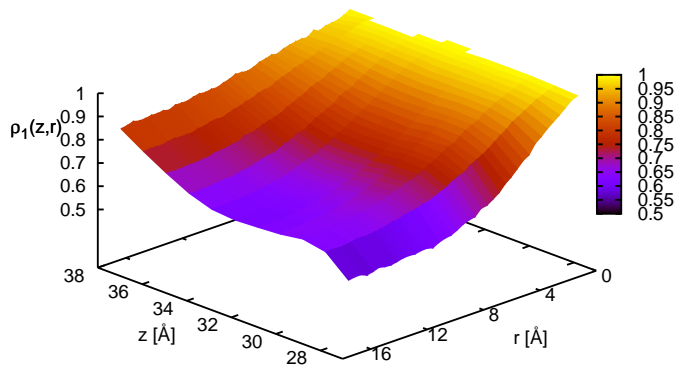}}%
\caption{(Color online) One-body 
distribution function $\rho_1(z,r)$ of \hd \ adsorbed on
$^4$He, with $r=\sqrt{x^2+y^2}$, at surface density $\sigma=0.0215$ \AA$^{-2}$. }  
\label{fig7}
\end{figure}

A key magnitude in the study of any quantum Bose gas is the one-body
distribution function $\rho_1(r)$ since it furnishes evidence of the presence of
off-diagonal long-range order in the system. As it is well known, its asymptotic behavior
in a homogeneous system $\lim_{r \rightarrow \infty} \rho_1(r)=n_0$   gives
the fraction of particles occupying the zero-momentum state, that is the
condensate fraction $n_0$. In Fig. \ref{fig7}, we show a surface plot
containing results of $\rho_1(z,r)$ at density  $\sigma=0.0215$ \AA$^{-2}$
obtained following the same method as in the grid of $g(z,r)$ shown in Fig.
\ref{fig5}. In the outer part of the density profile the condensate
fraction approaches one because the density is very small. When $z$
decreases the condensate fraction also decreases and reaches a plateau in
the central part of $\rho_{\text H}(r)$. If $z$ is reduced even more and
$\rho_{\text{He}}(r)$ starts to increase, the
\hd \  condensate fraction decreases again due to the small but nonzero $^4$He
density; the low statistics in this part makes the signal very noisy and
therefore we do not plot data for $z < 27$ \AA\ in Fig. \ref{fig7}.

\begin{figure}
\centerline{
\includegraphics[width=0.85\linewidth]{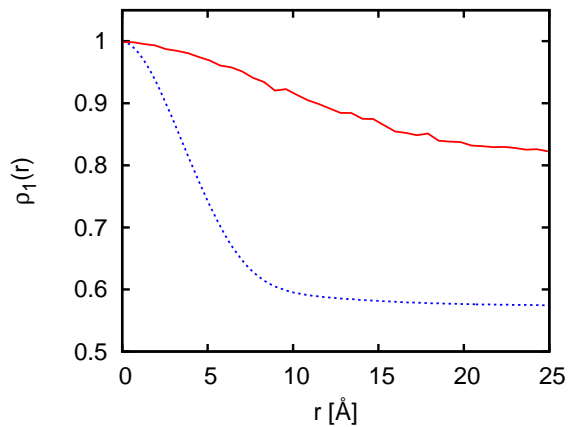}}%
\caption{(Color online) Comparison between the one-body distribution function in the
center of the slab, corresponding to a density $\sigma=0.0095$ \AA$^{-2}$ (solid
line) with the one corresponding to a purely 2D \hd \ gas at the same surface density
(dotted line).  }  
\label{fig8}
\end{figure}

A relevant issue in the study of the off-diagonal long-range order in the
adsorbed gas is the \textit{dimensionality} of the results achieved. As we
have made before for the two-body distribution functions, we compare
$\rho_1(r)$ for a 2D gas and for a slice in the center of the adsorbed
layer at the same density in Fig. \ref{fig8}. The results show that in this
case the behavior in the layer is significantly different from the one
observed in strictly 2D. The difference is larger than the one we have
observed at the same density for $g(r)$ (Fig. \ref{fig6}), with values for
the condensate fraction that differs in  $\sim 30$ \%. The condensate
fraction of the 2D gas is clearly smaller than the one of the layer due to
the transverse degree of freedom $z$ that translates into an effective surface
density smaller than the one of the full layer.

\begin{figure}
\centerline{
\includegraphics[width=0.85\linewidth]{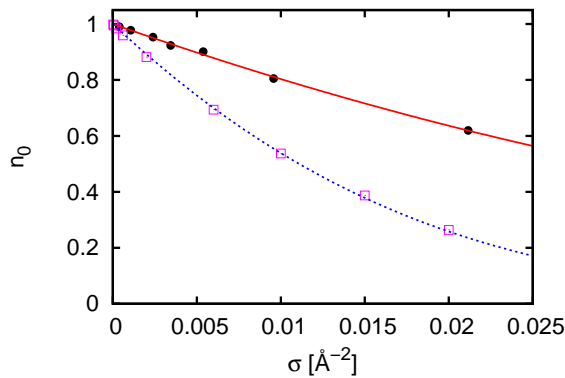}}%
\caption{(Color online) Condensate fraction as a function of the surface density
$\sigma$. Solid circles correspond to \hd \ on $^4$He and open squares to a 2D gas.
The lines on top of the DMC data are fits to guide the eye. }  
\label{fig9}
\end{figure}

The density dependence of the condensate fraction of adsorbed \hd \ is
shown in Fig. \ref{fig9}. The values reported have been obtained from the
asymptotic value of the one-body distribution function in the
central part of the density profile. As expected, the condensate fraction
is nearly 1 at very low densities and then decreases when $\sigma$
increases. However, the decrease is quite slow in such a way that even at
densities as large as $\sigma=0.02$ \AA$^{-2}$ the condensate fraction is
still $n_0 \simeq 0.6$. At the same density, the condensate fraction of the
2D gas is half this value, $n_0 \simeq 0.3$. The dependence of $n_0$ with
the density for the 2D geometry, shown in Fig. \ref{fig9} for comparison,
is significantly stronger with a larger depletion of the condensate
fraction for all densities.

\section{Summary and Conclusions}

The experimental realization of an extremely thin layer of \hd \ adsorbed
on the surface of superfluid $^4$He provides a unique opportunity for the
study of nearly two-dimensional quantum gases. The system is stable and the
influence of the liquid substrate is nearly negligible, without the
corrugation effects that a solid surface like graphite provides. Moreover,
spin-polarized hydrogen is a specially appealing system from the
theoretical side because it is the
best example of quantum matter (it remains gas even in the zero temperature
limit) and its interatomic interaction is known with high accuracy.    
In the present work, we have addressed its study from a microscopic
approach relying on the use of quantum Monte Carlo methods by means of a
simulation that incorporates the full Hamiltonian of the system, composed by
a realistic $^4$He surface and the  layer of \hd \ adsorbed on it.

From very low coverages up to relatively high surface densities, we have
reported results of the main properties of adsorbed \hd : energy, density
profile, two- and one-body distribution functions, and the condensate
fraction. Our results point to a $\sim 8$ \AA\  thick layer that virtually
\textit{floats} on top of $^4$He. We
have calculated the energy as a function of the surface density $\sigma$
and compared these energies with the results
obtained in a purely 2D \hd \ gas in order to establish the degree of
two-dimensionality of the layer.
The agreement between both simulations is
only satisfactory for small densities $\sigma \alt 5 \times 10^{-3}$
\AA$^{-2}$ and, from then on, the additional degree of freedom in the $z$
direction of the layer causes its energy to grow slower than in strictly
2D. Significant departures of strictly 2D behavior are also observed in the
two-body radial distribution function and mainly in the condensate fraction
values. Our DMC results show that the condensate fraction for the layer is
appreciably higher than in 2D, with values as large as $n_0=0.6$ at the
largest coverages studied. If we convert this coverage to volume density by using the layer width of 8 \AA, we see that the condensate fraction is quite close to published 3D values in Ref. \onlinecite{leandra1}. 
From these results we can be certain that a BKT phase transition
would be a realistic scenario at low surface densities. For higher
densities, further study using intensive
path-integral Monte Carlo simulations at finite temperatures would be
needed.

\begin{acknowledgments}
The authors acknowledge partial financial support from the  
DGI (Spain) Grant No.~FIS2011-25275, Generalitat de Catalunya 
Grant No.~2009SGR-1003, Qatar National 
Research Fund NPRP 5-674-1-114 as well as the support from MSES (Croatia)
 under Grant No.~177-1770508-0493.
\end{acknowledgments}

\end{document}